\documentclass{article}

\PassOptionsToPackage{numbers, compress}{natbib}
\usepackage[final]{neurips_2023_ml4ps}

\usepackage[utf8]{inputenc} 
\usepackage[T1]{fontenc}    
\usepackage{hyperref}       
\usepackage{url}            
\usepackage{booktabs}       
\usepackage{amsfonts}       
\usepackage{nicefrac}       
\usepackage{microtype}      
\usepackage{graphicx}
\usepackage{xcolor}         

\title{A Physics-Informed Variational Autoencoder for Rapid Galaxy Inference and Anomaly Detection}

\author{%
  Alexander~Gagliano \\
  The NSF AI Institute for Artificial Intelligence and Fundamental Interactions \\
  \texttt{gaglian2@mit.edu} \\
   \And
   V. Ashley Villar \\
  Center for Astrophysics \textbar{} Harvard \& Smithsonian\\
}

\begin{document}

\maketitle

\begin{abstract}
 The Vera C. Rubin Observatory is slated to observe nearly 20 billion galaxies during its decade-long Legacy Survey of Space and Time. The rich imaging data it collects will be an invaluable resource for probing galaxy evolution across cosmic time, characterizing the host galaxies of transient phenomena, and identifying novel populations of anomalous systems. To facilitate these studies, we introduce a convolutional variational autoencoder trained to estimate the redshift, stellar mass, and star-formation rates of galaxies from multi-band imaging data. We train and test our physics-informed CVAE on a spectroscopic sample of $\sim$26,000 galaxies within $z<1$ imaged through the Dark Energy Camera Legacy Survey. We show that our model can infer redshift and stellar mass more accurately than the latest image-based self-supervised learning approaches, and is >100x faster than more computationally-intensive SED-fitting techniques. Using a small sample of Green Pea and Red Spiral galaxies reported in the literature, we further demonstrate how this CVAE can be used to rapidly identify rare galaxy populations and interpret what makes them unique.
\end{abstract}

\section{Introduction}
At the cornerstone of a range of diverse astrophysical domains sits the chemical and dynamical histories of galaxies. Aspects of these histories can be inferred piecewise from observations. From broad-band photometry, one can infer the underlying spectral energy distribution (SED); using morphological features, one can reconstruct a galaxy's merger/interaction history. Novel galaxy sub-populations have been discovered through detailed manual inspection along these axes; some prominent examples include `Green Pea galaxies', defined by their compact shapes and high star-formation rates \cite{2009Cardamone_GreenPeas}; and `passive Red Spirals'-- spiral galaxies with minimal star formation, whose colors seem to contradict a morphology that is commonly tied to much more active galaxies \cite{2010Masters_PassiveReds}. Several of these sub-populations have been only identified through manual vetting by scores of human volunteers (e.g., \cite{2009Schawinski_GalaxyZoo}).

The rise of deep convolutional neural networks, which allow for morphological classification of galaxies from image data \cite{2021Cavanagh_CNNs}, and of simulation-based inference to estimate their underlying SED parameters from photometry \cite{2023Wang_SBI}, have paved the way for low-latency galaxy analysis in the era of synoptic surveys. However, anomaly detection within these frameworks remains an open question: out-of-distribution events may be identified by flagging high reconstruction errors at test time \cite[e.g.,][]{torabi2023practical}, but this approach can produce samples contaminated by non-astrophysical anomalies such as poorly-calibrated images and chip gaps. 

Here, we introduce a technique to guide the latent features of a convolutional variational auto-encoder (CVAE) toward the physical parameters of a galaxy. CVAEs are a class of neural network designed for compression. Variational AEs, in particular, typically employ the Kullback–Leibler (KL) divergence to enforce a continuous and probabilistic latent distribution characterized by a multivariate standard Normal. CVAEs have been extensively used in the literature for galaxy image reconstruction \cite{2016Ravanbakhsh_VAE, 2021Cheng_VAE, 2021Arcelin_VAE}. Here, we extend this methodology to disentangle the observational properties of a galaxy (e.g., its orientation) from its intrinsic properties in the latent space, and discover more meaningful anomalies. 

\section{Data and Methodology}
In this section, we describe our methodology for designing and training a physics-informed CVAE. Our CVAE encodes a galaxy image into 5 latent dimensions, the first four corresponding to the orientation, redshift, stellar mass, and star-formation rate of the galaxy. The fifth dimension encodes any remaining structure not captured by the first four features.

We select a sample of $z<1$ spectroscopic galaxies from both the LSST deep drilling fields catalogue from \cite{2022Zou_SEDs} (hereafter, the Z22 sample) and from GAMA DR4 \cite{2022Driver_GAMA}. These catalogues report the average star-formation rate and stellar mass of each galaxy obtained with the SED-fitting codes \texttt{CIGALE} and \texttt{PROSPECT}, respectively, along with their associated uncertainties. We include the Z22 sample to provide training support at high redshift, and the GAMA catalogue to includes low-redshift objects for which rich morphological information is available. For galaxies that overlap with the Dark Energy Camera Legacy Survey (DECaLS) footprint, we then download 128x128 px DECaLS co-added images centered at the location of each galaxy in $grz$\footnote{\url{https://www.legacysurvey.org/decamls/}}. DECaLS data reaches a 5-$\sigma$ depth of $g\sim$24, making it a sufficient analog for upcoming Rubin data ($g\sim$24.5\footnote{\url{https://www.lsst.org/scientists/keynumbers}}). We remove five galaxies that are missing images in at least one filter, leaving us with a combined sample of 26,126 galaxies.

As in \cite{2021Arcelin_GalaxyImageScaling}, we pre-process our images in each filter $b$ by applying the following normalization: 
\begin{equation}
x_b = \rm{tanh}\Biggl(\rm{sinh}^{-1}\Big(\beta \frac{x_{\rm{raw,b}}}{\langle max(x_{\rm{raw}, b})\rangle_b}\Big)\Biggr)
\end{equation}

where $\beta = 10$ was chosen empirically to allow our image maxima to be well-distributed within $[0, 1]$. We then interpolate each image to 69x69 px to reduce training times.

Next, we use the \texttt{photutils} package to estimate the orientation of each galaxy. We stack the $grz$ images and subtract the median background level from the stack, assuming a 10x10 px box size and a 3x3 px median filter. We then convolve each background-subtracted image with a 2D Gaussian filter and construct a catalog of all identified sources. The morphological features of each source are calculated by default, and we extract the orientation of the source whose centroid is closest to the center of the image. We then split our data into an 80\% training sample and a 20\% validation sample.


To assess the utility of our CVAE for anomaly detection, we augment our validation set with a sample of 294 Red Spiral galaxies from \cite{2010Masters_PassiveReds} and the KISS sample of 13 Green Pea galaxies \cite{2022KISS}. The Red Spiral galaxies were consolidated by the 160,000 volunteers of the Galaxy Zoo project. 

We construct our CVAE using the same architecture as in \cite{WuArchitecture}\footnote{\url{https://github.com/jwuphysics/galaxy-autoencoders.}}. The network consists of three convolution encoding layers, three deconvolution decoding layers, and a latent dimension of 5. We train the network using a loss function consisting of three terms. The first is the reconstruction loss for an image of $n$ pixels, the second is the KL divergence between the fifth latent feature and a standard univariate Normal, and the third is the mean squared error between the means of the first four latent distributions $\mu$ and the galaxy's physical properties:

\begin{equation}
    L = \beta_0 \frac{1}{n} \sum_{i=1}^n (x_i - \hat{x_i})^2 +  ( \sigma_5^2 + \mu_5^2 - \mathrm{log}(\sigma_5) - 1) +  \beta_1 \sum_{k=1}^4 \frac{1}{\sigma_{p_k}^2} (p_k - \mu_k)^2 
\end{equation}

where $p$ is a vector containing the physical properties of the galaxy (orientation, spectroscopic redshift, and survey-reported stellar mass and star-formation rate) and $\sigma_{p}$ are their reported 1-$\sigma$ uncertainties (we assume an uncertainty of 0.1 rad for all calculated orientations). We leave the fifth latent dimension of our CVAE free to capture the remaining morphological information in each image. The coefficients $\beta_0$ and $\beta_1$ are hyperparameters; we have found through moderate tuning that setting $\beta_0 = 10$ and $\beta_1 = 50$ results in reasonable parameter estimates without significantly degrading the quality of reconstructed images. 

As a baseline comparison, we train a vanilla CVAE with the same architecture but excluding the physical term from the loss function (the KL-divergence is now calculated across all 5 latent features, which are unphysical). We train each network for 1000 epochs using the standard \verb|Adam| optimizer \cite{2014AdamOptimizer} with a learning rate of $5\times10^{-6}$ and a batch size of 128 on two Nvidia A100 80GB GPUs, confirming convergence of the loss and saving the model weights at the epoch where the validation loss is minimized. Our code and trained models can be found at the associated Github repository for this work\footnote{\url{https://github.com/alexandergagliano/physics_CVAE}}.

\section{Results and Discussion}



\begin{figure}
\centering
\includegraphics[width=\linewidth]{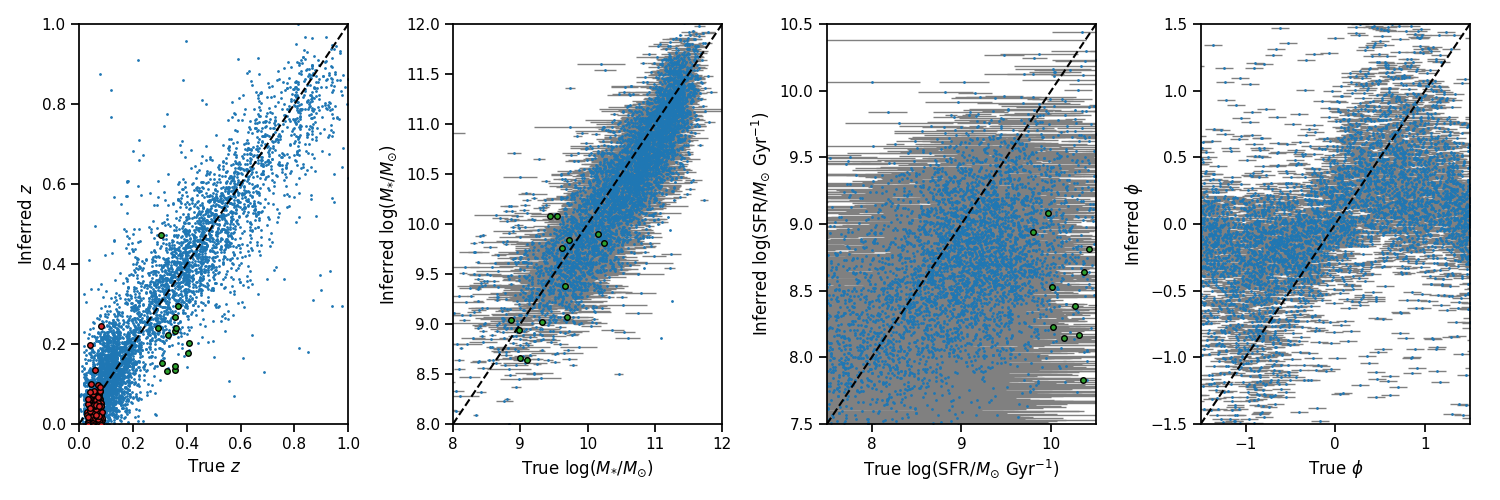}\label{fig:LearnedParameters}
  \caption{A comparison between the learned latent features of the CVAE applied to the validation set, and the physical properties of the associated galaxies derived through SED fitting. 1-$\sigma$ uncertainties on SED-estimated parameters are shown in gray. Literature values for Green Pea galaxies are given in green, and those for Red Spirals are given in red (only redshift is reported for Red Spirals).}
\end{figure}

\begin{table}
\begin{center}
\begin{tabular}{ccc} \toprule
    {Method} & {Redshift $R^2$} & {Stellar Mass $R^2$} \\ \midrule
    $(r,g,z)$ Photometry + MLP \cite{2023Lanusse_AstroCLIP}  & 0.69 & 0.65 \\ \midrule
   Image Embedding + MLP  \cite{2021Stein_Embedding} & 0.39  & 0.45   \\ \midrule
    Image Embedding + MLP \cite{2023Lanusse_AstroCLIP}  & 0.63   & 0.57   \\  \midrule
    Image Embedding + kNN \cite{2023Lanusse_AstroCLIP}   & 0.71   & 0.66   \\  \midrule 
   \textbf{Image Embedding (ours)}  & \textbf{0.83}   & \textbf{0.75}   \\ 
   \bottomrule
\end{tabular}
\caption{$R^2$ values for the linear regression models that predict a galaxy's spectroscopically-derived redshift and stellar mass from its corresponding CVAE latent feature.}
\end{center}
\end{table}

The catalog-level and CVAE-estimated physical properties for the validation sample are compared in Fig.~\ref{fig:LearnedParameters}. The CVAE appears to successfully recover the physical parameters of galaxies in the validation set. The predictions are unbiased, with mean absolute deviations of at most 0.1, for star-formation rate. In addition, few catastrophic outliers are observed for most galaxy properties (the scatter is large for star-formation rate, but comparable to the error bars on publicly reported SED estimates). An exception is the orientation $\phi$: because the parameter is periodic in nature (galaxies with orientations $\pi/2$ and $-\pi/2$ are indistinguishable), this feature is poorly reconstructed at high inclinations. 

To quantitatively evaluate the CVAE's performance on the validation set, we fit a linear model between each CVAE-inferred feature and its spectroscopically-derived counterpart. We report the $R^2$ value of these regression models for spectroscopic redshift and stellar mass in Table 1, alongside those reported for other image-based parameter inference methods from \cite{2023Lanusse_AstroCLIP}. The values from \cite{2023Lanusse_AstroCLIP} were calculated on a sample of $\sim$200k $z<0.6$ DECaLS galaxies, roughly an order of magnitude greater than the sample considered in this work. Despite the smaller sample available for training, our technique far outperforms the reported techniques (particularly those trained using self-supervised learning) for estimating these galaxy properties, and to higher redshifts ($z<1$). Our $R^2$ value for star-formation rate is 0.26; though unavoidable degeneracies exist at the photometric level, tighter constraints from SED fits would likely improve the CVAE's ability to infer this property in the future. 


Surprisingly, our network is able to predict reasonable estimates for the spectroscopic redshifts and stellar masses of the Red Spiral and Green Pea galaxies from images alone, despite their anomalous colors. The redshifts of the Green Peas are still systematically underestimated by the model, which skews the predicted star-formation rates downward. 

\begin{figure}
\centering
\includegraphics[width=0.9\linewidth]{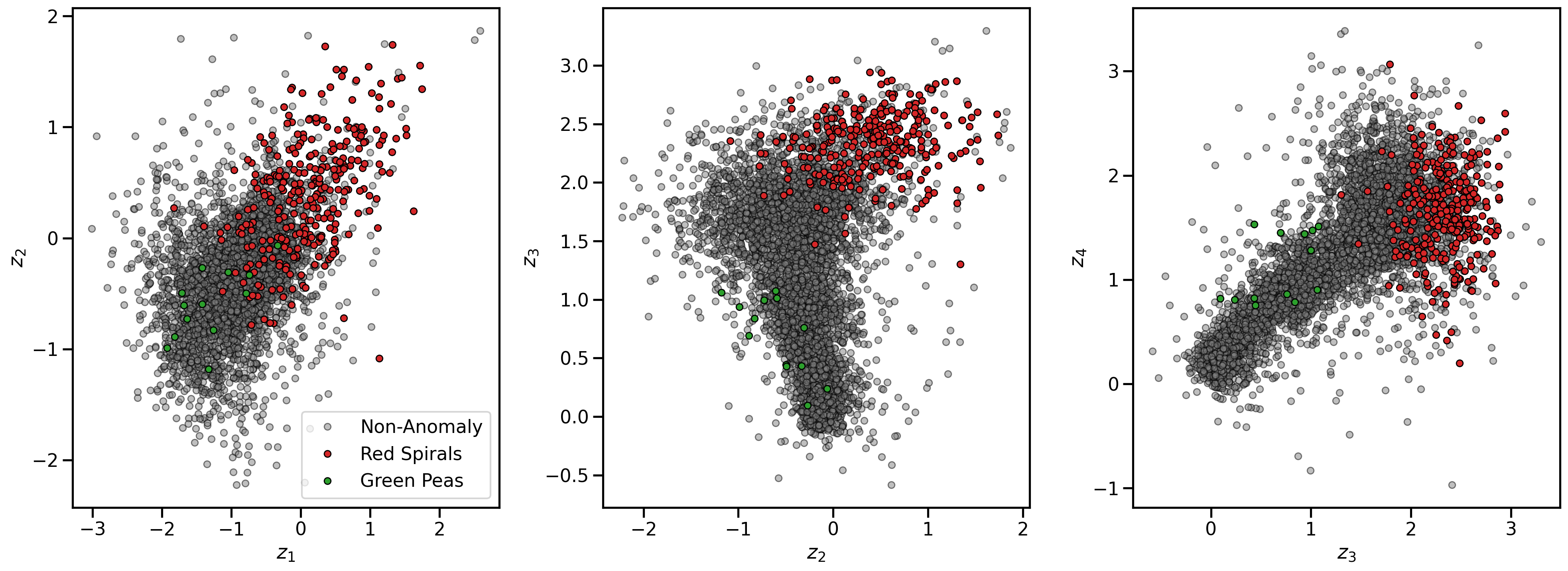}\label{fig:Anomaly_Uninformed}
\includegraphics[width=0.9\linewidth]{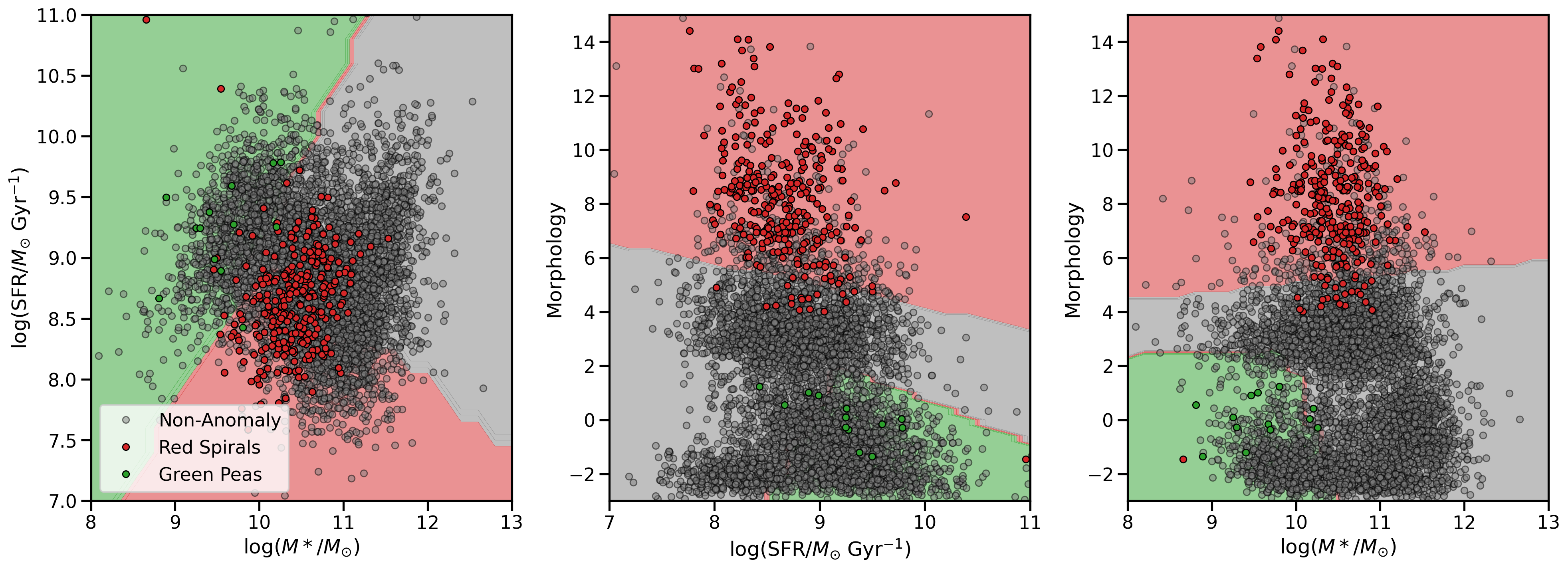}\label{fig:Anomaly_PhysicsInformed}
  \caption{Projections of the latent features drawn for the validation set, including the two populations of anomalies. Sub-types are given in legend. \textbf{Top row:} Latent space for the vanilla CVAE, trained without physics-informed loss constraints. \textbf{Bottom Row:} Latent space for our physics-informed CVAE, with two-dimensional SVM decision boundaries overplotted. Green Peas are identified by their compact morphologies, low stellar masses, and high star-formation rates; Red Spirals are identified by their extended morphologies, higher stellar masses, and lower star-formation rates.}
\end{figure}

We next construct a support vector machine (SVM) classifier to recover our sample of anomalous galaxies using only encoded features. We use the implementation from \texttt{scikit-learn}, with a radial basis function (RBF) kernel and scaled kernel coefficients. In each of five trials, we randomly split the combined sample into 80\% train and 20\% test sets. We then balance our three classes across the training set with \texttt{SMOTE} and evaluate performance on the test set. We calculate the fraction of each class recovered by the classifier in each case, and average the model's performance across all five trials. We find that our CVAE recovers $94\pm2\%$ of Red Spiral galaxies with a $5\%$ contamination rate from non-anomalous galaxies, and $67\pm0\%$ of Green Pea galaxies with a $2\%$ contamination rate from non-anomalous galaxies (though we caution that each test sample only contains 2-3 Green Peas). The same classifier trained on the latent features of the vanilla CVAE recovers $94\pm2\%$ of Red Spiral galaxies with a $6\%$ contamination rate and $47\pm27\%$ of Green Pea galaxies with a $4\%$ contamination rate. We conclude that the physics-informed CVAE achieves marginally superior, and significantly more consistent, performance over the vanilla CVAE in recovering Green Peas. 


We next plot 2D projections of the latent space for the physics-informed and vanilla CVAE models in Fig.~\ref{fig:Anomaly_PhysicsInformed}. For the physics-informed plots, we include the decision boundaries from a 2d SVM with linear kernel. Though the Red Spirals fall toward the edges of the latent parameter distributions in the baseline, the plots lack physical intuition. The latent space of the physics-informed CVAE, however, reveals the properties that make these sub-populations anomalous. Green Pea galaxies push toward the highest star-formation rates, lowest stellar masses, and lowest ``morphology'' values of the sample (this can be considered a proxy for compactness). Red Spirals, in contrast, are the most extended objects in the sample and have low star-formation rates for their stellar masses, with most falling in or below the `green valley' that distinguishes active from quenched galaxies \cite{2014GreenValley}. 

\begin{figure}
\centering
\includegraphics[width=\linewidth]{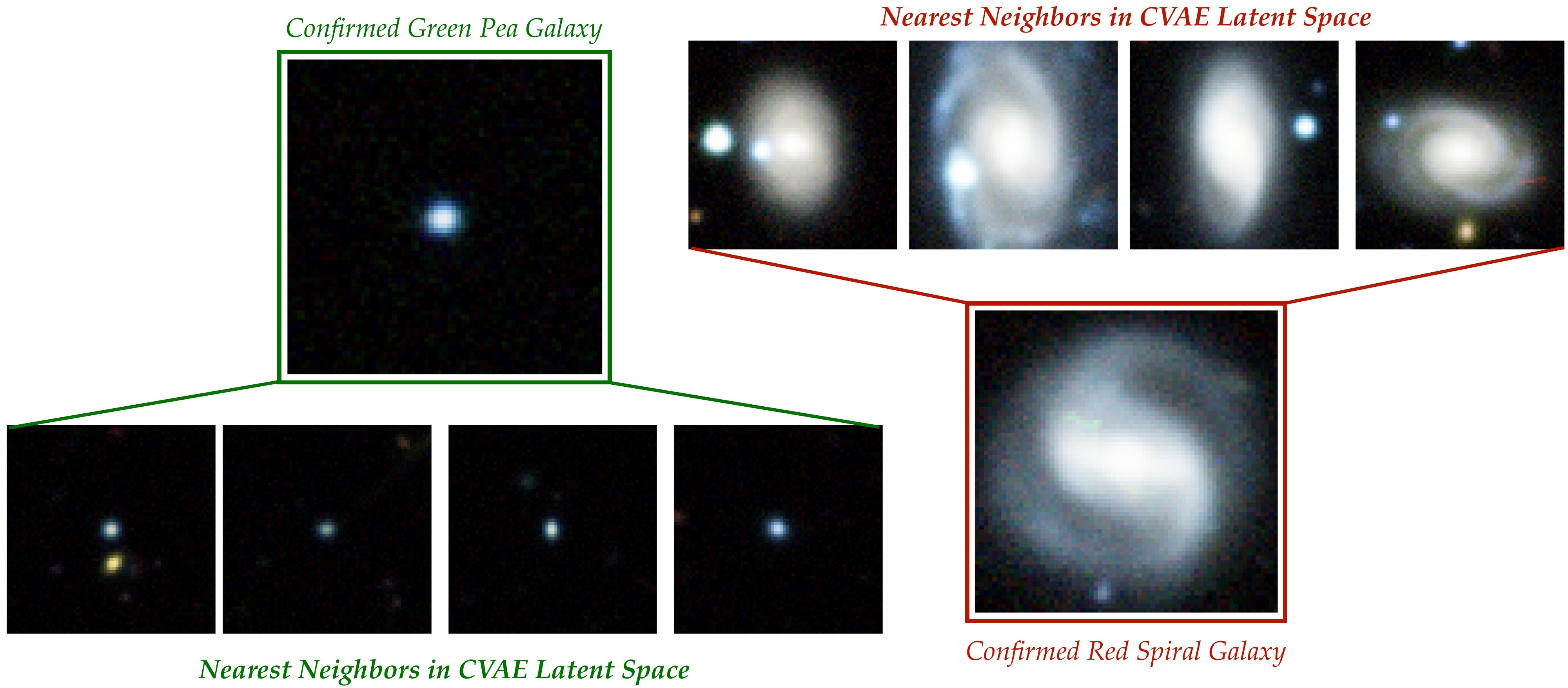}\label{fig:AnomalyLookAlikes}
  \caption{Latent space nearest-neighbors for the anomalies manually flagged in the Galaxy Zoo project. \textbf{Left:} Images of galaxies with latent properties most similar to labeled ``Green Pea'' galaxies. \textbf{Right:} Images of galaxies most similar to labeled ``Red Spiral'' galaxies. These galaxies are less blue than the majority of other spirals in our sample, suggesting older average stellar populations.}
\end{figure}

Finally, we identify galaxies in our sample most similar to the labeled Green Peas and Red Spirals. We use the \texttt{NearestNeighbor} method in \texttt{scikit-learn} to identify ten galaxies in the non-anomalous sample with latent features most similar to each galaxy in the anomalous sample. We present a sample of galaxies with lowest latent-space Minkowski distance to multiple Red Spirals, and those with lowest distance to multiple Green Peas, in Fig.~\ref{fig:AnomalyLookAlikes}. These galaxies are prime targets for upcoming follow-up studies, to confirm their associations and further characterize their evolutionary histories.


The ability to rapidly infer physical parameters competitive with state-of-the-art SED models using only three images of a galaxy is encouraging. Once trained, evaluation of our CVAE takes $7$ ms per galaxy on an Apple M2 Max chip. This is substantially less than the $\sim$10 hrs required for rigorous SED fitting using modern approaches \cite{2021Johnson_Prospector}, and $>100$x faster than existing simulation-based inference (SBI) models optimized for this task \cite{2023Wang_SBI}. With our approach, estimating the redshift, stellar mass, and star-formation rate of all 20B galaxies observed by the Vera C. Rubin Observatory would require only $\sim$1500 CPU-hours (albeit without the added benefit of full posterior distributions, as are provided by SBI). New, neural network-based inference approaches like the one presented here will help bring scalability and interpretability to petabyte-scale data exploration among astronomical surveys, and facilitate the search for new galaxy sub-types.

\acksection
We thank the anonymous reviewers for helpful feedback which improved this manuscript. This work is supported by the National Science Foundation under Cooperative Agreement PHY-2019786 (The NSF AI Institute for Artificial Intelligence and Fundamental Interactions, http://iaifi.org/). VAV acknowledges support by the NSF through grant AST-2108676.

The Legacy Surveys consist of three individual and complementary projects: the Dark Energy Camera Legacy Survey (DECaLS; Proposal ID \#2014B-0404; PIs: David Schlegel and Arjun Dey), the Beijing-Arizona Sky Survey (BASS; NOAO Prop. ID \#2015A-0801; PIs: Zhou Xu and Xiaohui Fan), and the Mayall z-band Legacy Survey (MzLS; Prop. ID \#2016A-0453; PI: Arjun Dey). DECaLS, BASS and MzLS together include data obtained, respectively, at the Blanco telescope, Cerro Tololo Inter-American Observatory, NSF’s NOIRLab; the Bok telescope, Steward Observatory, University of Arizona; and the Mayall telescope, Kitt Peak National Observatory, NOIRLab. Pipeline processing and analyses of the data were supported by NOIRLab and the Lawrence Berkeley National Laboratory (LBNL). The Legacy Surveys project is honored to be permitted to conduct astronomical research on Iolkam Du’ag (Kitt Peak), a mountain with particular significance to the Tohono O’odham Nation.

NOIRLab is operated by the Association of Universities for Research in Astronomy (AURA) under a cooperative agreement with the National Science Foundation. LBNL is managed by the Regents of the University of California under contract to the U.S. Department of Energy.

This project used data obtained with the Dark Energy Camera (DECam), which was constructed by the Dark Energy Survey (DES) collaboration. Funding for the DES Projects has been provided by the U.S. Department of Energy, the U.S. National Science Foundation, the Ministry of Science and Education of Spain, the Science and Technology Facilities Council of the United Kingdom, the Higher Education Funding Council for England, the National Center for Supercomputing Applications at the University of Illinois at Urbana-Champaign, the Kavli Institute of Cosmological Physics at the University of Chicago, Center for Cosmology and Astro-Particle Physics at the Ohio State University, the Mitchell Institute for Fundamental Physics and Astronomy at Texas A\&M University, Financiadora de Estudos e Projetos, Fundacao Carlos Chagas Filho de Amparo, Financiadora de Estudos e Projetos, Fundacao Carlos Chagas Filho de Amparo a Pesquisa do Estado do Rio de Janeiro, Conselho Nacional de Desenvolvimento Cientifico e Tecnologico and the Ministerio da Ciencia, Tecnologia e Inovacao, the Deutsche Forschungsgemeinschaft and the Collaborating Institutions in the Dark Energy Survey. The Collaborating Institutions are Argonne National Laboratory, the University of California at Santa Cruz, the University of Cambridge, Centro de Investigaciones Energeticas, Medioambientales y Tecnologicas-Madrid, the University of Chicago, University College London, the DES-Brazil Consortium, the University of Edinburgh, the Eidgenossische Technische Hochschule (ETH) Zurich, Fermi National Accelerator Laboratory, the University of Illinois at Urbana-Champaign, the Institut de Ciencies de l’Espai (IEEC/CSIC), the Institut de Fisica d’Altes Energies, Lawrence Berkeley National Laboratory, the Ludwig Maximilians Universitat Munchen and the associated Excellence Cluster Universe, the University of Michigan, NSF’s NOIRLab, the University of Nottingham, the Ohio State University, the University of Pennsylvania, the University of Portsmouth, SLAC National Accelerator Laboratory, Stanford University, the University of Sussex, and Texas A\&M University.

BASS is a key project of the Telescope Access Program (TAP), which has been funded by the National Astronomical Observatories of China, the Chinese Academy of Sciences (the Strategic Priority Research Program “The Emergence of Cosmological Structures” Grant \#XDB09000000), and the Special Fund for Astronomy from the Ministry of Finance. The BASS is also supported by the External Cooperation Program of Chinese Academy of Sciences (Grant \#114A11KYSB20160057), and Chinese National Natural Science Foundation (Grant \#12120101003, \#11433005).

The Legacy Survey team makes use of data products from the Near-Earth Object Wide-field Infrared Survey Explorer (NEOWISE), which is a project of the Jet Propulsion Laboratory/California Institute of Technology. NEOWISE is funded by the National Aeronautics and Space Administration.

The Legacy Surveys imaging of the DESI footprint is supported by the Director, Office of Science, Office of High Energy Physics of the U.S. Department of Energy under Contract No. DE-AC02-05CH1123, by the National Energy Research Scientific Computing Center, a DOE Office of Science User Facility under the same contract; and by the U.S. National Science Foundation, Division of Astronomical Sciences under Contract No. AST-0950945 to NOAO.

GAMA is a joint European-Australasian project based around a spectroscopic campaign using the Anglo-Australian Telescope. The GAMA input catalogue is based on data taken from the Sloan Digital Sky Survey and the UKIRT Infrared Deep Sky Survey. Complementary imaging of the GAMA regions is being obtained by a number of independent survey programmes including GALEX MIS, VST KiDS, VISTA VIKING, WISE, Herschel-ATLAS, GMRT and ASKAP providing UV to radio coverage. GAMA is funded by the STFC (UK), the ARC (Australia), the AAO, and the participating institutions. The GAMA website is http://www.gama-survey.org/.






\medskip

{
\small

\bibliography{neurips}

\begin{thebibliography}{19}
\providecommand{\natexlab}[1]{#1}
\providecommand{\url}[1]{\texttt{#1}}
\expandafter\ifx\csname urlstyle\endcsname\relax
  \providecommand{\doi}[1]{doi: #1}\else
  \providecommand{\doi}{doi: \begingroup \urlstyle{rm}\Url}\fi

\bibitem[WuA()]{WuArchitecture}
What do galaxies look like? learning from variational autoencoders.
\newblock \url{https://jwuphysics.substack.com/p/galaxy-autoencoders}.
\newblock Accessed: 2023-09-29.

\bibitem[{Arcelin} et~al.(2021{\natexlab{a}}){Arcelin}, {Doux}, {Aubourg}, {Roucelle}, and {LSST Dark Energy Science Collaboration}]{2021Arcelin_GalaxyImageScaling}
B.~{Arcelin}, C.~{Doux}, E.~{Aubourg}, C.~{Roucelle}, and {LSST Dark Energy Science Collaboration}.
\newblock {Deblending galaxies with variational autoencoders: A joint multiband, multi-instrument approach}.
\newblock \emph{Monthly Notices of the RAS}, 500\penalty0 (1):\penalty0 531--547, Jan. 2021{\natexlab{a}}.
\newblock \doi{10.1093/mnras/staa3062}.

\bibitem[{Arcelin} et~al.(2021{\natexlab{b}}){Arcelin}, {Doux}, {Aubourg}, {Roucelle}, and {LSST Dark Energy Science Collaboration}]{2021Arcelin_VAE}
B.~{Arcelin}, C.~{Doux}, E.~{Aubourg}, C.~{Roucelle}, and {LSST Dark Energy Science Collaboration}.
\newblock {Deblending galaxies with variational autoencoders: A joint multiband, multi-instrument approach}.
\newblock \emph{Monthly Notices of the RAS}, 500\penalty0 (1):\penalty0 531--547, Jan. 2021{\natexlab{b}}.
\newblock \doi{10.1093/mnras/staa3062}.

\bibitem[{Brunker} et~al.(2022){Brunker}, {Salzer}, {Kimsey-Miller}, and {Cousins}]{2022KISS}
S.~W. {Brunker}, J.~J. {Salzer}, B.~{Kimsey-Miller}, and B.~{Cousins}.
\newblock {The Environments of Green Pea Galaxies. I. The KISS Sample}.
\newblock \emph{Astrophysical Journal}, 926\penalty0 (2):\penalty0 131, Feb. 2022.
\newblock \doi{10.3847/1538-4357/ac469f}.

\bibitem[{Cardamone} et~al.(2009){Cardamone}, {Schawinski}, {Sarzi}, {Bamford}, {Bennert}, {Urry}, {Lintott}, {Keel}, {Parejko}, {Nichol}, {Thomas}, {Andreescu}, {Murray}, {Raddick}, {Slosar}, {Szalay}, and {Vandenberg}]{2009Cardamone_GreenPeas}
C.~{Cardamone}, K.~{Schawinski}, M.~{Sarzi}, S.~P. {Bamford}, N.~{Bennert}, C.~M. {Urry}, C.~{Lintott}, W.~C. {Keel}, J.~{Parejko}, R.~C. {Nichol}, D.~{Thomas}, D.~{Andreescu}, P.~{Murray}, M.~J. {Raddick}, A.~{Slosar}, A.~{Szalay}, and J.~{Vandenberg}.
\newblock {Galaxy Zoo Green Peas: discovery of a class of compact extremely star-forming galaxies}.
\newblock \emph{Monthly Notices of the RAS}, 399\penalty0 (3):\penalty0 1191--1205, Nov. 2009.
\newblock \doi{10.1111/j.1365-2966.2009.15383.x}.

\bibitem[{Cavanagh} et~al.(2021){Cavanagh}, {Bekki}, and {Groves}]{2021Cavanagh_CNNs}
M.~K. {Cavanagh}, K.~{Bekki}, and B.~A. {Groves}.
\newblock {Morphological classification of galaxies with deep learning: comparing 3-way and 4-way CNNs}.
\newblock \emph{Monthly Notices of the RAS}, 506\penalty0 (1):\penalty0 659--676, Sept. 2021.
\newblock \doi{10.1093/mnras/stab1552}.

\bibitem[{Cheng} et~al.(2021){Cheng}, {Huertas-Company}, {Conselice}, {Arag{\'o}n-Salamanca}, {Robertson}, and {Ramachandra}]{2021Cheng_VAE}
T.-Y. {Cheng}, M.~{Huertas-Company}, C.~J. {Conselice}, A.~{Arag{\'o}n-Salamanca}, B.~E. {Robertson}, and N.~{Ramachandra}.
\newblock {Beyond the hubble sequence - exploring galaxy morphology with unsupervised machine learning}.
\newblock \emph{Monthly Notices of the RAS}, 503\penalty0 (3):\penalty0 4446--4465, May 2021.
\newblock \doi{10.1093/mnras/stab734}.

\bibitem[{Driver} et~al.(2022){Driver}, {Bellstedt}, {Robotham}, {Baldry}, {Davies}, {Liske}, {Obreschkow}, {Taylor}, {Wright}, {Alpaslan}, {Bamford}, {Bauer}, {Bland-Hawthorn}, {Bilicki}, {Bravo}, {Brough}, {Casura}, {Cluver}, {Colless}, {Conselice}, {Croom}, {de Jong}, {D'Eugenio}, {De Propris}, {Dogruel}, {Drinkwater}, {Dvornik}, {Farrow}, {Frenk}, {Giblin}, {Graham}, {Grootes}, {Gunawardhana}, {Hashemizadeh}, {H{\"a}u{\ss}ler}, {Heymans}, {Hildebrandt}, {Holwerda}, {Hopkins}, {Jarrett}, {Heath Jones}, {Kelvin}, {Koushan}, {Kuijken}, {Lara-L{\'o}pez}, {Lange}, {L{\'o}pez-S{\'a}nchez}, {Loveday}, {Mahajan}, {Meyer}, {Moffett}, {Napolitano}, {Norberg}, {Owers}, {Radovich}, {Raouf}, {Peacock}, {Phillipps}, {Pimbblet}, {Popescu}, {Said}, {Sansom}, {Seibert}, {Sutherland}, {Thorne}, {Tuffs}, {Turner}, {van der Wel}, {van Kampen}, and {Wilkins}]{2022Driver_GAMA}
S.~P. {Driver}, S.~{Bellstedt}, A.~S.~G. {Robotham}, I.~K. {Baldry}, L.~J. {Davies}, J.~{Liske}, D.~{Obreschkow}, E.~N. {Taylor}, A.~H. {Wright}, M.~{Alpaslan}, S.~P. {Bamford}, A.~E. {Bauer}, J.~{Bland-Hawthorn}, M.~{Bilicki}, M.~{Bravo}, S.~{Brough}, S.~{Casura}, M.~E. {Cluver}, M.~{Colless}, C.~J. {Conselice}, S.~M. {Croom}, J.~{de Jong}, F.~{D'Eugenio}, R.~{De Propris}, B.~{Dogruel}, M.~J. {Drinkwater}, A.~{Dvornik}, D.~J. {Farrow}, C.~S. {Frenk}, B.~{Giblin}, A.~W. {Graham}, M.~W. {Grootes}, M.~L.~P. {Gunawardhana}, A.~{Hashemizadeh}, B.~{H{\"a}u{\ss}ler}, C.~{Heymans}, H.~{Hildebrandt}, B.~W. {Holwerda}, A.~M. {Hopkins}, T.~H. {Jarrett}, D.~{Heath Jones}, L.~S. {Kelvin}, S.~{Koushan}, K.~{Kuijken}, M.~A. {Lara-L{\'o}pez}, R.~{Lange}, {\'A}.~R. {L{\'o}pez-S{\'a}nchez}, J.~{Loveday}, S.~{Mahajan}, M.~{Meyer}, A.~J. {Moffett}, N.~R. {Napolitano}, P.~{Norberg}, M.~S. {Owers}, M.~{Radovich}, M.~{Raouf}, J.~A. {Peacock}, S.~{Phillipps}, K.~A. {Pimbblet}, C.~{Popescu}, K.~{Said}, A.~E. {Sansom}, M.~{Seibert},
  W.~J. {Sutherland}, J.~E. {Thorne}, R.~J. {Tuffs}, R.~{Turner}, A.~{van der Wel}, E.~{van Kampen}, and S.~M. {Wilkins}.
\newblock {Galaxy And Mass Assembly (GAMA): Data Release 4 and the z < 0.1 total and z < 0.08 morphological galaxy stellar mass functions}.
\newblock \emph{Monthly Notices of the RAS}, 513\penalty0 (1):\penalty0 439--467, June 2022.
\newblock \doi{10.1093/mnras/stac472}.

\bibitem[{Johnson} et~al.(2021){Johnson}, {Leja}, {Conroy}, and {Speagle}]{2021Johnson_Prospector}
B.~D. {Johnson}, J.~{Leja}, C.~{Conroy}, and J.~S. {Speagle}.
\newblock {Stellar Population Inference with Prospector}.
\newblock \emph{Astrophysical Journal, Supplement}, 254\penalty0 (2):\penalty0 22, June 2021.
\newblock \doi{10.3847/1538-4365/abef67}.

\bibitem[{Kingma} and {Ba}(2014)]{2014AdamOptimizer}
D.~P. {Kingma} and J.~{Ba}.
\newblock {Adam: A Method for Stochastic Optimization}.
\newblock \emph{arXiv e-prints}, art. arXiv:1412.6980, Dec. 2014.
\newblock \doi{10.48550/arXiv.1412.6980}.

\bibitem[{Lanusse} et~al.(2023){Lanusse}, {Parker}, {Golkar}, {Cranmer}, {Bietti}, {Eickenberg}, {Krawezik}, {McCabe}, {Ohana}, {Pettee}, {Regaldo-Saint Blancard}, {Tesileanu}, {Cho}, and {Ho}]{2023Lanusse_AstroCLIP}
F.~{Lanusse}, L.~{Parker}, S.~{Golkar}, M.~{Cranmer}, A.~{Bietti}, M.~{Eickenberg}, G.~{Krawezik}, M.~{McCabe}, R.~{Ohana}, M.~{Pettee}, B.~{Regaldo-Saint Blancard}, T.~{Tesileanu}, K.~{Cho}, and S.~{Ho}.
\newblock {AstroCLIP: Cross-Modal Pre-Training for Astronomical Foundation Models}.
\newblock \emph{arXiv e-prints}, art. arXiv:2310.03024, Oct. 2023.
\newblock \doi{10.48550/arXiv.2310.03024}.

\bibitem[{Masters} et~al.(2010){Masters}, {Mosleh}, {Romer}, {Nichol}, {Bamford}, {Schawinski}, {Lintott}, {Andreescu}, {Campbell}, {Crowcroft}, {Doyle}, {Edmondson}, {Murray}, {Raddick}, {Slosar}, {Szalay}, and {Vandenberg}]{2010Masters_PassiveReds}
K.~L. {Masters}, M.~{Mosleh}, A.~K. {Romer}, R.~C. {Nichol}, S.~P. {Bamford}, K.~{Schawinski}, C.~J. {Lintott}, D.~{Andreescu}, H.~C. {Campbell}, B.~{Crowcroft}, I.~{Doyle}, E.~M. {Edmondson}, P.~{Murray}, M.~J. {Raddick}, A.~{Slosar}, A.~S. {Szalay}, and J.~{Vandenberg}.
\newblock {Galaxy Zoo: passive red spirals}.
\newblock \emph{Monthly Notices of the RAS}, 405\penalty0 (2):\penalty0 783--799, June 2010.
\newblock \doi{10.1111/j.1365-2966.2010.16503.x}.

\bibitem[{Ravanbakhsh} et~al.(2016){Ravanbakhsh}, {Lanusse}, {Mandelbaum}, {Schneider}, and {Poczos}]{2016Ravanbakhsh_VAE}
S.~{Ravanbakhsh}, F.~{Lanusse}, R.~{Mandelbaum}, J.~{Schneider}, and B.~{Poczos}.
\newblock {Enabling Dark Energy Science with Deep Generative Models of Galaxy Images}.
\newblock \emph{arXiv e-prints}, art. arXiv:1609.05796, Sept. 2016.
\newblock \doi{10.48550/arXiv.1609.05796}.

\bibitem[{Salim}(2014)]{2014GreenValley}
S.~{Salim}.
\newblock {Green Valley Galaxies}.
\newblock \emph{Serbian Astronomical Journal}, 189:\penalty0 1--14, Dec. 2014.
\newblock \doi{10.2298/SAJ1489001S}.

\bibitem[{Schawinski} et~al.(2009){Schawinski}, {Lintott}, {Thomas}, {Sarzi}, {Andreescu}, {Bamford}, {Kaviraj}, {Khochfar}, {Land}, {Murray}, {Nichol}, {Raddick}, {Slosar}, {Szalay}, {VandenBerg}, and {Yi}]{2009Schawinski_GalaxyZoo}
K.~{Schawinski}, C.~{Lintott}, D.~{Thomas}, M.~{Sarzi}, D.~{Andreescu}, S.~P. {Bamford}, S.~{Kaviraj}, S.~{Khochfar}, K.~{Land}, P.~{Murray}, R.~C. {Nichol}, M.~J. {Raddick}, A.~{Slosar}, A.~{Szalay}, J.~{VandenBerg}, and S.~K. {Yi}.
\newblock {Galaxy Zoo: a sample of blue early-type galaxies at low redshift*}.
\newblock \emph{Monthly Notices of the RAS}, 396\penalty0 (2):\penalty0 818--829, June 2009.
\newblock \doi{10.1111/j.1365-2966.2009.14793.x}.

\bibitem[{Stein} et~al.(2021){Stein}, {Harrington}, {Blaum}, {Medan}, and {Lukic}]{2021Stein_Embedding}
G.~{Stein}, P.~{Harrington}, J.~{Blaum}, T.~{Medan}, and Z.~{Lukic}.
\newblock {Self-supervised similarity search for large scientific datasets}.
\newblock \emph{arXiv e-prints}, art. arXiv:2110.13151, Oct. 2021.
\newblock \doi{10.48550/arXiv.2110.13151}.

\bibitem[Torabi et~al.(2023)Torabi, Mirtaheri, and Greco]{torabi2023practical}
H.~Torabi, S.~L. Mirtaheri, and S.~Greco.
\newblock Practical autoencoder based anomaly detection by using vector reconstruction error.
\newblock \emph{Cybersecurity}, 6\penalty0 (1):\penalty0 1, 2023.

\bibitem[{Wang} et~al.(2023){Wang}, {Leja}, {Villar}, and {Speagle}]{2023Wang_SBI}
B.~{Wang}, J.~{Leja}, V.~A. {Villar}, and J.~S. {Speagle}.
\newblock {SBI$^{++}$: Flexible, Ultra-fast Likelihood-free Inference Customized for Astronomical Applications}.
\newblock \emph{Astrophysical Journal, Letters}, 952\penalty0 (1):\penalty0 L10, July 2023.
\newblock \doi{10.3847/2041-8213/ace361}.

\bibitem[{Zou} et~al.(2022){Zou}, {Brandt}, {Chen}, {Leja}, {Ni}, {Yan}, {Yang}, {Zhu}, {Luo}, {Nyland}, {Vito}, and {Xue}]{2022Zou_SEDs}
F.~{Zou}, W.~N. {Brandt}, C.-T. {Chen}, J.~{Leja}, Q.~{Ni}, W.~{Yan}, G.~{Yang}, S.~{Zhu}, B.~{Luo}, K.~{Nyland}, F.~{Vito}, and Y.~{Xue}.
\newblock {Spectral Energy Distributions in Three Deep-drilling Fields of the Vera C. Rubin Observatory Legacy Survey of Space and Time: Source Classification and Galaxy Properties}.
\newblock \emph{Astrophysical Journal, Supplement}, 262\penalty0 (1):\penalty0 15, Sept. 2022.
\newblock \doi{10.3847/1538-4365/ac7bdf}.

\end{thebibliography}


}


\end{document}